 \documentclass[final,3p,times]{elsarticle}

\usepackage{graphics}
\usepackage{graphicx}
\usepackage{epsfig}
\usepackage{amssymb}

\journal{Physics Letters A}

\begin{document}
\begin{frontmatter}


\title{Magnetic response of a disordered binary ferromagnetic alloy to an oscillating magnetic
field}

\author[1,2]{Erol Vatansever\corref{cor1}}
\cortext[cor1]{Corresponding author. Tel.: +90 3019547; fax: +90 2324534188.} \ead{erol.vatansever@deu.edu.tr}
\author[1]{Hamza Polat}

\address[1]{Department of Physics, Dokuz Eyl\"{u}l University,
Tr-35160 \.{I}zmir, Turkey}

\address[2]{Dokuz Eyl\"{u}l University, Graduate School of Natural
and Applied Sciences, Turkey}


\begin{abstract}
By means of Monte Carlo simulation with local spin update Metropolis algorithm,
we have elucidated non-equilibrium phase transition properties and stationary-state treatment
of a disordered binary ferromagnetic alloy of the type $A_{p}B_{1-p}$ on a square lattice.
After a detailed analysis, we have found that the system shows many interesting and unusual
thermal and magnetic behaviors, for instance, the locations of dynamic phase transition points
change significantly depending upon amplitude and period of the external magnetic field as well
as upon the active  concentration of $A-$ type components. Much effort has also been dedicated to
clarify the hysteresis tools, such as coercivity, dynamic loop area as well as dynamic
correlations between time  dependent magnetizations and external time dependent
applied field as a functions of period and amplitude of field as well as active
concentration of of $A-$ type components,   and outstanding physical findings have been reported in
order to better understand the dynamic process underlying present system.
\end{abstract}

\end{frontmatter}
\section{Introduction}\label{introduction}
\hspace{0.4cm} It was shown, for the first time, by  theoretically
that when a ferromagnetic material with coupling $J$ is exposed to a time dependent
driving oscillating  magnetic field, the system may not respond to the external magnetic
field instantaneously, which gives rise to the existence of unusual and interesting
behaviors due to the competing time scales of the relaxation behavior
of the system  and period of the external magnetic field \cite{Tome}. A typical ferromagnet
exists in dynamically  disordered (\textbf{P}) phase where the time dependent magnetization
oscillates around value of zero for the high temperature and amplitude of field regimes.
At this stage, the time dependent magnetization of system is capable  to follow the external
field with relatively small phase lag. However,  it oscillates around a non-zero value which indicates
a dynamically ordered (\textbf{F}) phase for low temperatures and small amplitude of field regimes.
After that, a great deal of studies concerning the non-equilibrium phase transitions and also
hysteresis behaviors of different types of magnetic systems has been investigated by using both
experimental and theoretical techniques. For the sake of completeness,
it is beneficial to talk about some of the prominent experimental and theoretical
studies, respectively. Actually,  most of the experimental works have been devoted to
the dynamic phase transitions and hysteretic treatments of different types of thin
films \cite{Suen, He, Jiang, Choi, Robb, Berger}. For example, in Ref. \cite{Choi},
dynamic magnetization reversal behavior  of polycrystalline $\mathrm{Ni_{80}Fe_{20}}$ films has
been studied by applying a magnetic field along the easy axis of the sample and
it is found that the hysteresis loop  area $A\left(=\oint MdH\right)$  is found to follow the
scaling relation $A\propto H_{0}^{\alpha}\Omega^{\beta}T^{-\gamma}$ with $\alpha
\approx 0.9$, $\beta \approx0.8$ and $\gamma =0.38$, where $H_{0}, \; \Omega$ and $T$ are the amplitude, frequency of the
external magnetic field and temperature, respectively. Moreover, it has been shown recently
by Berger  et al. in Ref. \cite{Berger} that uniaxial $\mathrm{Co}$  film sample under the influence of
both bias (namely time independent) and  time dependent  oscillating magnetic fields in
the vicinity of dynamic phase transition displays  transient  behavior for $\tau<\tau_{c}$,
where $\tau$ and $\tau_{c}$ are period and critical period of the external applied field.
Based on the experimental investigations we mentioned briefly above, it has been discovered
that experimental non-equilibrium dynamics of  considered real magnetic systems strongly
resemble the dynamic behavior predicted from  theoretical calculations of
a kinetic Ising model.

On the other hand, from the theoretical point of view, dynamic phase  transitions as well as
hysteresis properties of both  bulk and finite size  lattice systems have been clarified
within the frameworks of Mean-Field  Theory (MFT) \cite{Acharyya1, Buendia, Punya, Keskin,
Keskin2,  Idigoras},  Effective-Field  Theory (EFT) with single-site correlations
\cite{Shi, Shi2, Yuksel, Akinci, Vatansever1, Vatansever2, Ozan,
Shi3}, as well as method of Monte Carlo (MC) simulations \cite{Shi3, Lo,  Chatterjee,
Park, Acharyya2, Acharyya3, Acharyya4, Acharyya5, Tauscher, Park2,
Yuksel2, Vatansever3, Vatansever4}.  For example, by benefiting from a detailed large-scale
MC simulation, in Ref. \cite{Park2},  Park and Pleimling considered kinetic
Ising models with surfaces  subjected to a periodic oscillating
magnetic field to probe the  role of surfaces at dynamic phase transitions.
They reported that the non-equilibrium surface universality class differs from that of the
equilibrium system, although the same universality class prevails for the corresponding
bulk systems. Moreover, by making use of MC simulation with local spin update Metropolis
scheme, dynamic phase transition features and stationary-state behavior of a
ferrimagnetic nanoparticle system with core-shell structure have been recently
analyzed by some of us in Ref. \cite{Yuksel2}. It has been observed that
the particle may exhibit a phase transition from \textbf{P} to \textbf{F}
phase with increasing ferromagnetic shell thickness in  the presence of
ultra-fast switching fields. Apart from these,  there exists a limited
number of   theoretical studies about the non-equilibrium  phase transition properties
of magnetic systems containing quenched  randomness resulting
from random interactions between the spins  with the same magnitudes or from a random
dilution of the   magnetic ions with non-magnetic  species on
the  magnetic  materials \cite{Vatansever1, Vatansever2, Zheng,
Ozan2,  Vatansever5}. It is clear that these studies
have a crucial role  to have a better insight of the physics
behind on real materials since a lot of magnetic materials have
some small defects, and magnetic properties, i.e. phase transition
temperature point of sample varies significantly depending on
the type of defects.

In this letter, for the first time we intend to  determine the magnetic phase
transition and hysteretic  properties of a disordered binary ferromagnetic alloy
of the type $A_{p}B_{1-p}$ system under a magnetic field that oscillates in time.
Such a quenched disordered system  consists of  two different species of
magnetic  components, namely $A$ and $B$,
and we select the magnetic  components $A$ and $B$ to
be as $S_{A}=1/2$  and $S_{B}=1$, respectively. Here,
$p$ refers to the concentration of type-$A$ magnetic ion. The square
lattice sites  are randomly occupied by $A$ and $B$ atoms depending
on the selected concentrations of magnetic components. Before  going further,
we should note that equilibrium or static features of  such of disordered
binary magnetic systems have been analyzed by means of several types of
frameworks such as  Perturbation Theory \cite{Thorpe}, MFT \cite{Thorpe, Tahir, Plascak, Katsura},
Bethe-Peirls Approximation \cite{Ishikawa}, EFT with single-site
correlations \cite{Honmura1, Kaneyoshi1, Kaneyoshi2, Kaneyoshi3, Kaneyoshi4,
Kaneyoshi5, Kaneyoshi6}  and method of MC  simulation \cite{Tatsumi, Scholten,
Scholten2, Godoy, Cambui}. For instance, the critical properties of
random mixtures of ferromagnetic and anti-ferromagnetic spin-spin interactions have been
studied with MC simulation on a simple cubic lattice in Ref. \cite{Tatsumi}.
It has been found that the  system exhibits spin-glass phase characterized by a cusp-like peak in
the susceptibility. Additionally, it is claimed by the author in Ref. \cite{Plascak}
that random-site binary ferromagnetic Ising model shows seven
topologically different types of phase diagrams including a variety
of multi-critical points within the framework MFT.

The main motivation of the present study is to look  answers for the physical facts
underlying below questions:

\begin{itemize}
 \item What is the effect of the amplitude and frequency of the
 external magnetic field on the dynamic phase transition
 properties (i.e. critical temperature) of the considered system?

\item What kind of physical relationships exist between the
 magnetic properties (i.e. dynamic loop area, dynamic
 correlation and coercivity) of the system and the concentration of $A$-type magnetic
 components?  In other words, whether  the macroscopic
treatments of the observable quantities of system depend on the concentration of $A$-type magnetic
 components or not.
\end{itemize}

The plan of the remainder parts of the paper is as follows:
In section (\ref{formulation}) we briefly present our
model. The results and discussions are presented in
section (\ref{discussion}), and finally  section (\ref{conclusion})
contains our conclusions.

\section{Formulation}\label{formulation}
\begin{figure}[!here]
\begin{center}
\includegraphics[width=3.15cm,height=3.15cm]{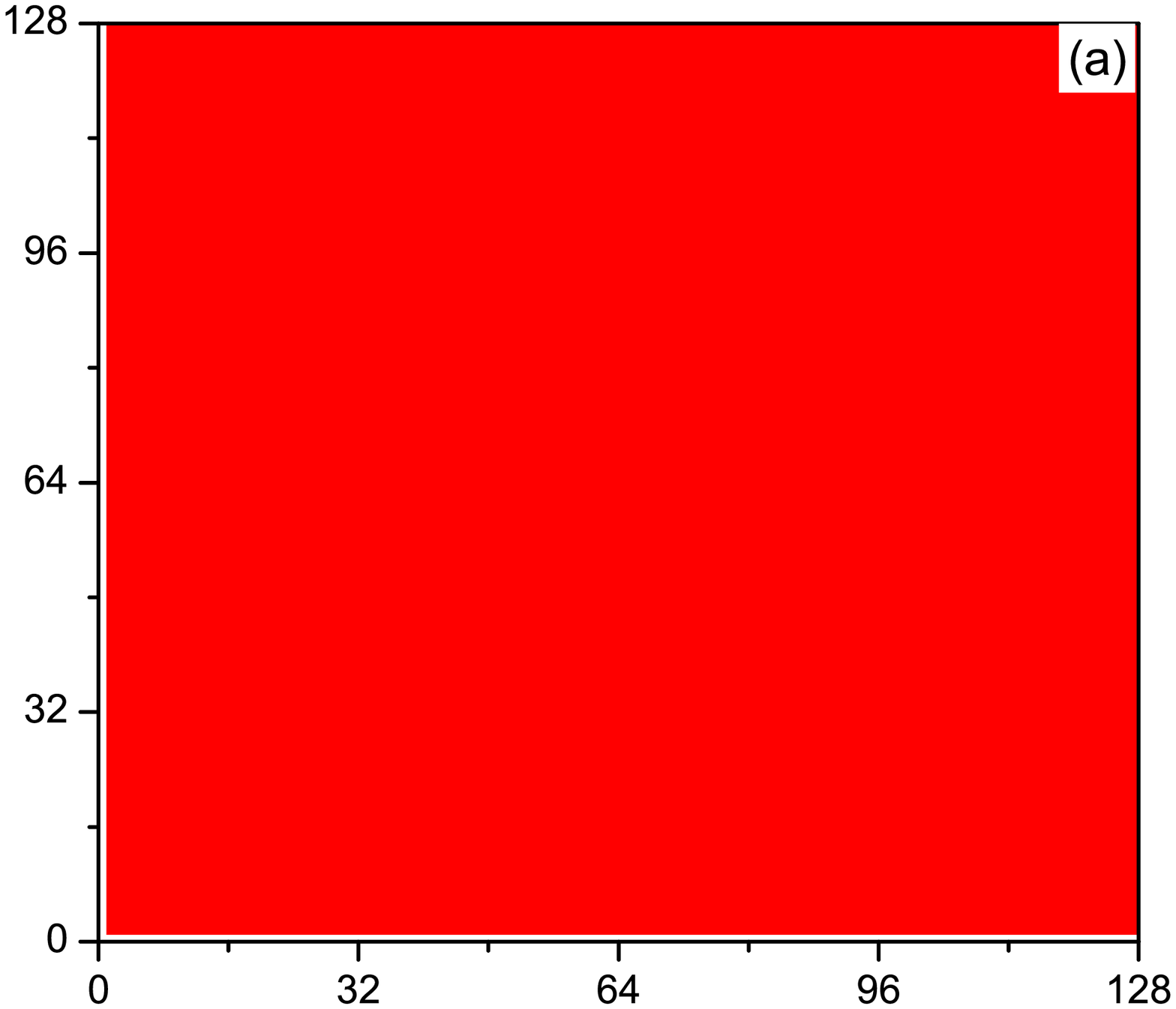}
\includegraphics[width=3.15cm,height=3.15cm]{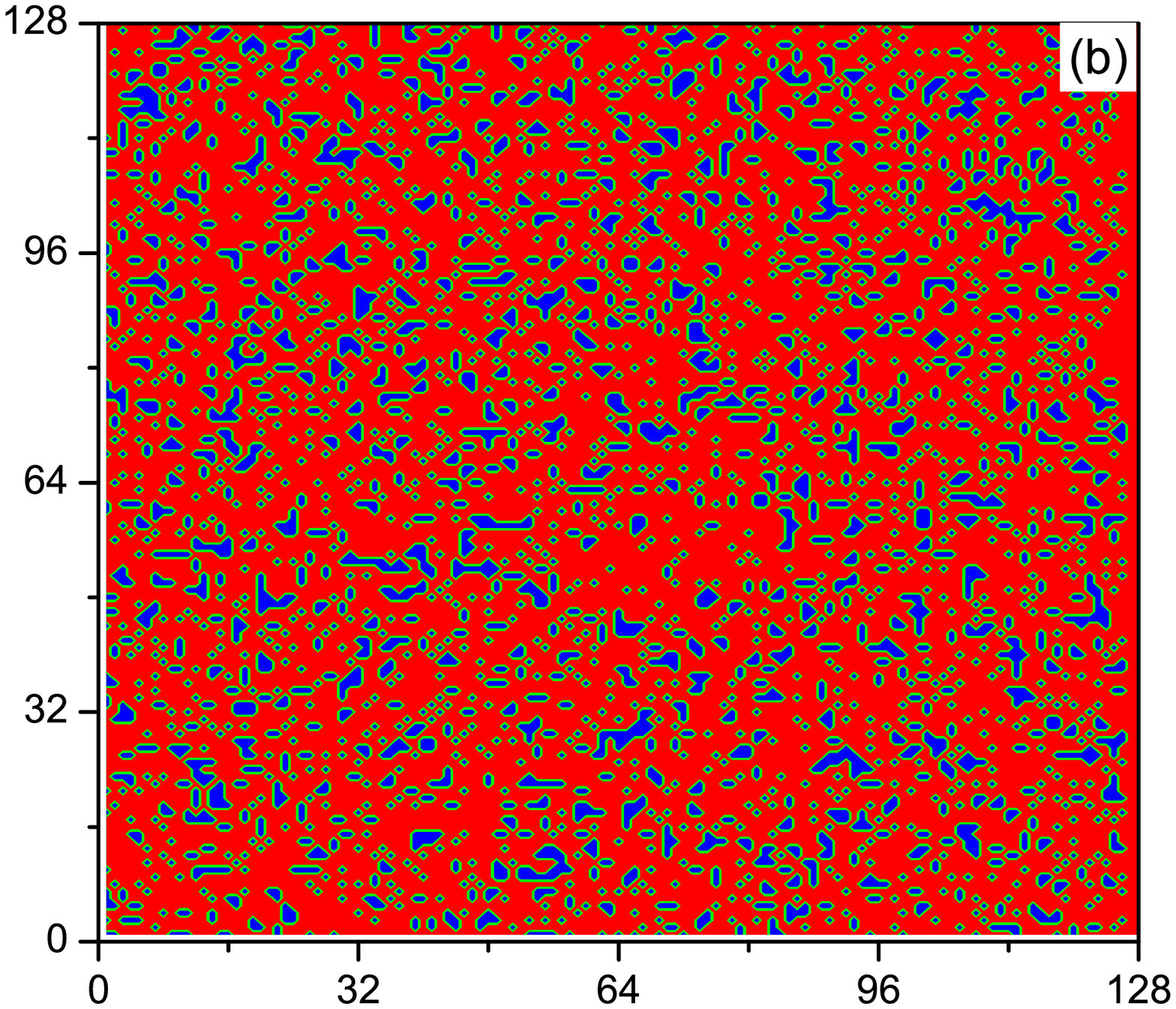}
\includegraphics[width=3.15cm,height=3.15cm]{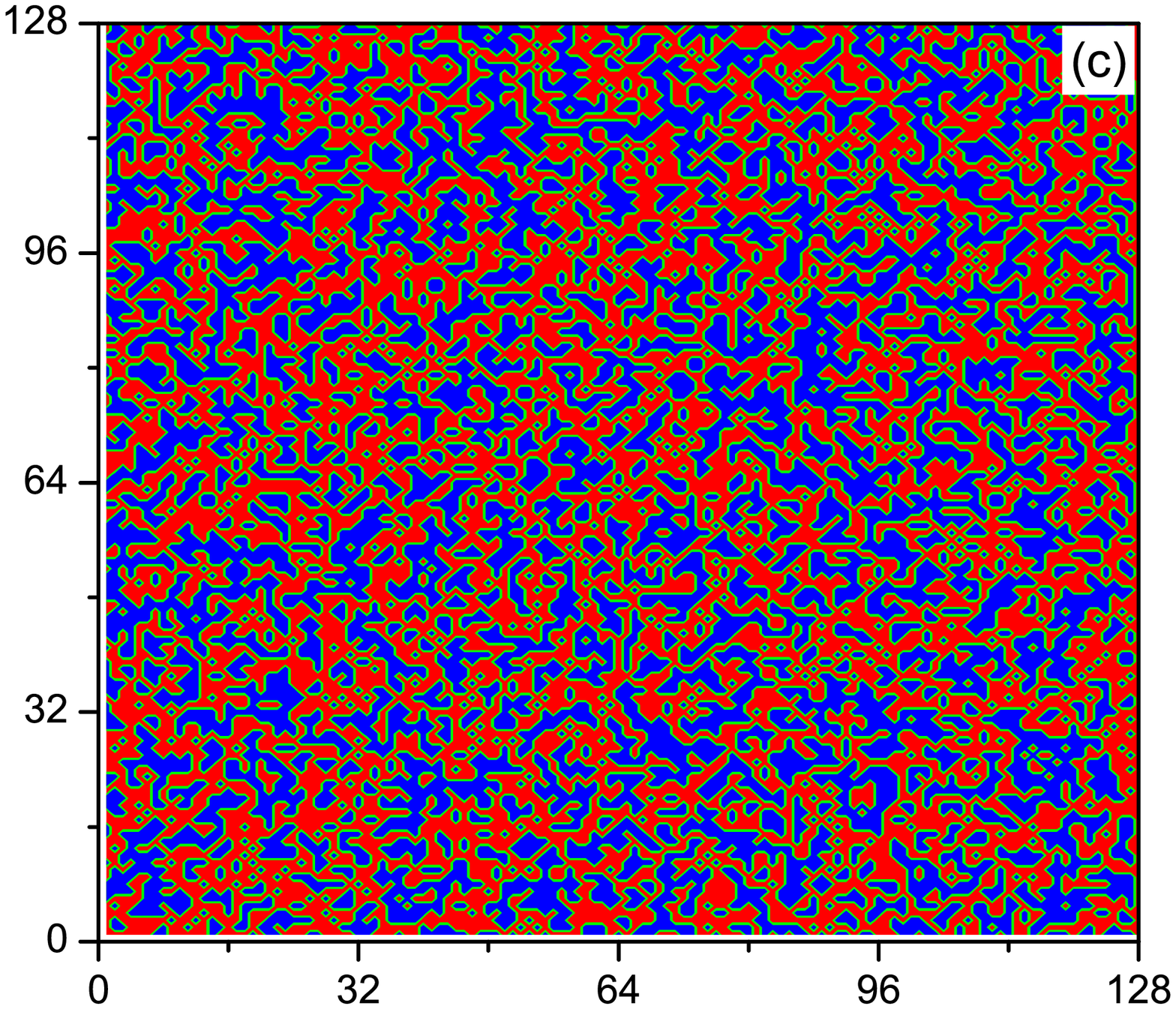}
\includegraphics[width=3.15cm,height=3.15cm]{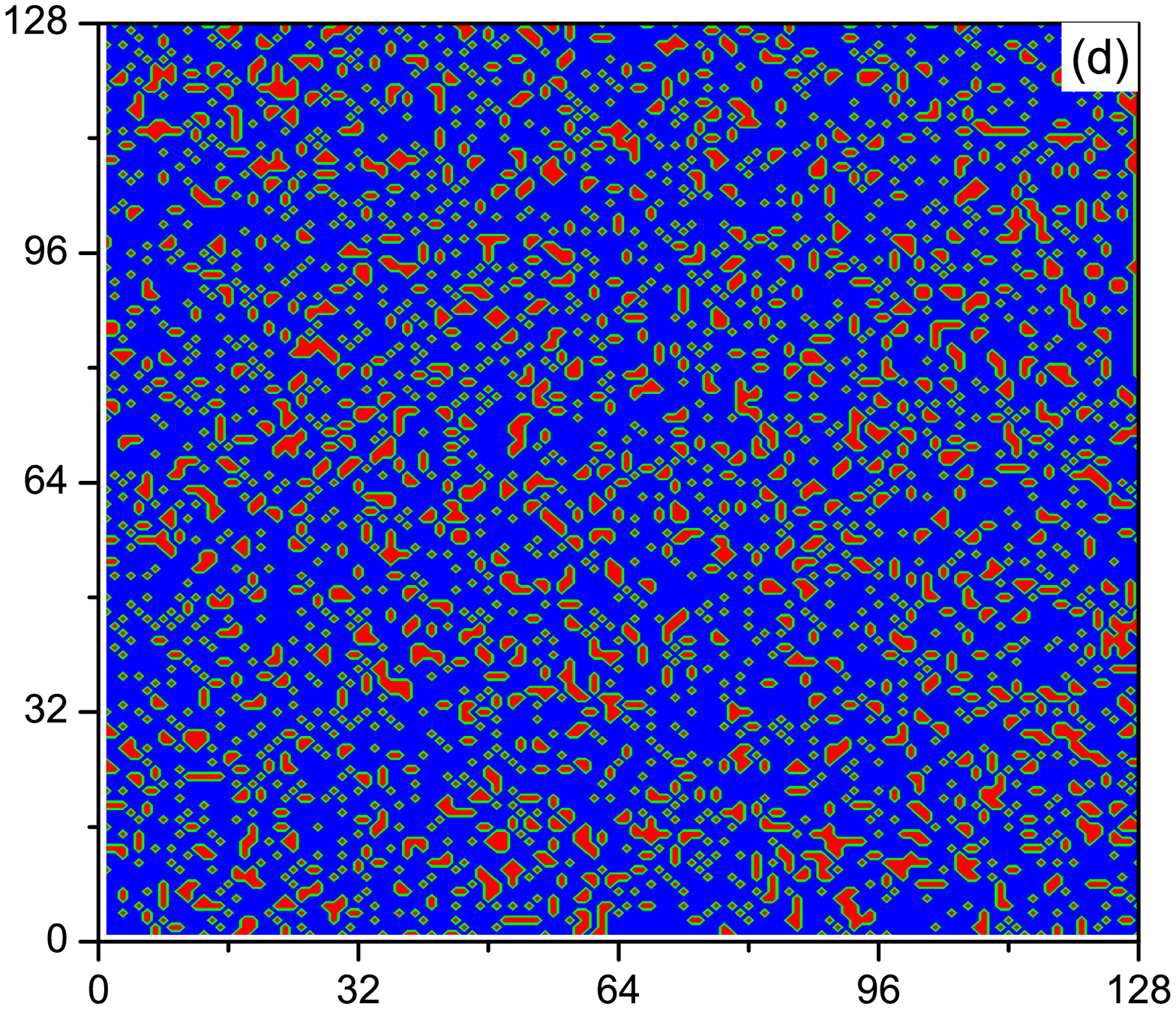}
\includegraphics[width=3.15cm,height=3.15cm]{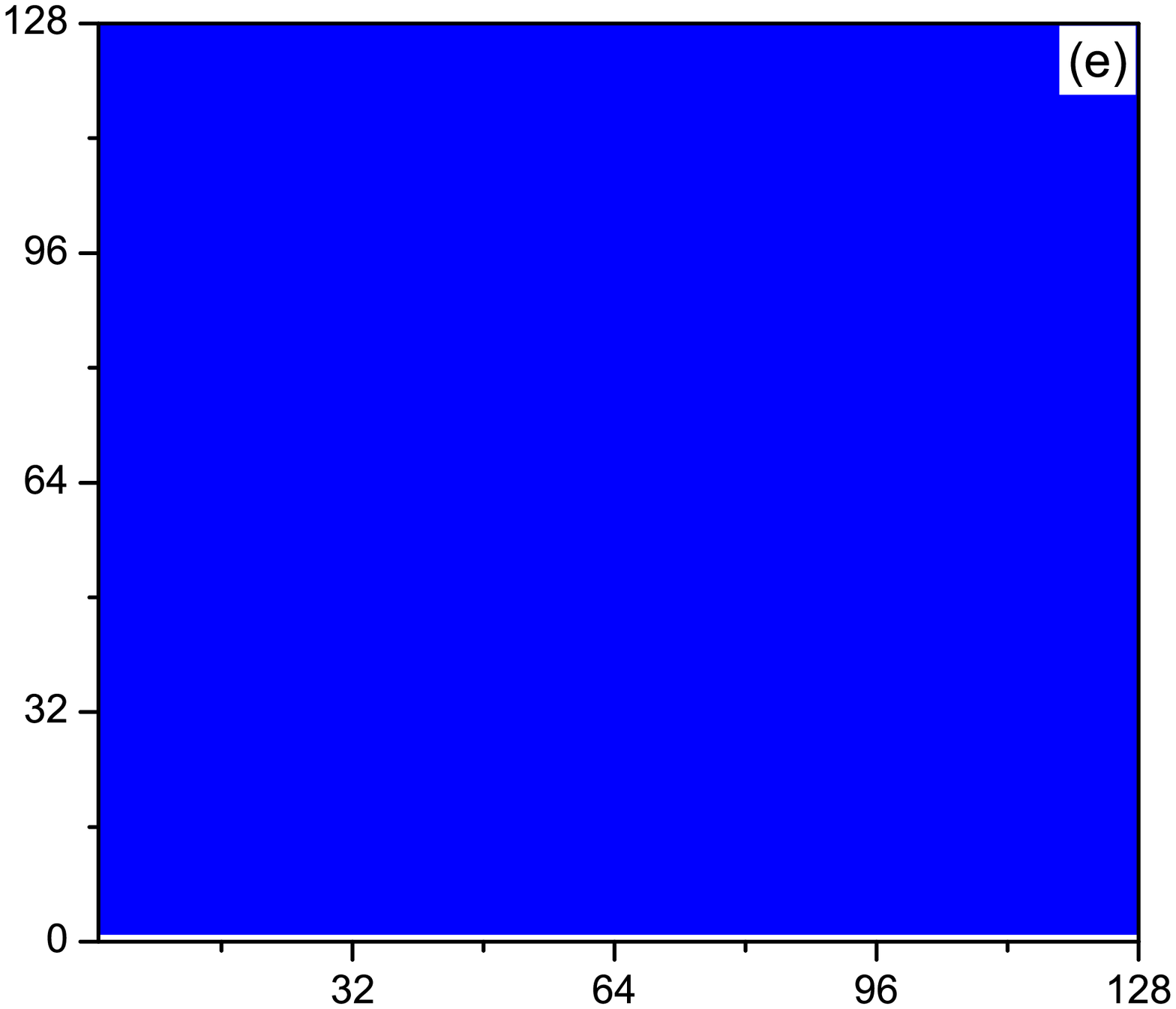}
\caption{Schematic representation of the system of the type $A_{p}B_{1-p}$
for five selected active concentration values of $A$-type magnetic component, namely (a) $p=0.0$,  (b) $p=0.25$,
(c) $p=0.5$, (d) $p=0.75$ and (e) $p=1.0$. The blue and red colors correspond to $A$ and $B$ types of
magnetic components which are randomly distributed.}\label{Fig1}
\end{center}
\end{figure}

The system simulated here, which is schematically shown in Fig. \ref{Fig1},
is a disordered binary ferromagnetic alloy of the type $A_{p}B_{1-p}$ which is
defined on a two dimensional regular  square lattice under the existence of a time
dependent driving oscillating magnetic field. The lattice sites are randomly occupied by two
different species of magnetic components $A$ and $B$ with the concentration $p$ and $1-p$, respectively.
Thus, the Hamiltonian of the considered system can be written in the following form:
\begin{equation}\label{eq1}
\hat{H}=-J\sum_{\langle i,j \rangle}\left[\delta_{i}\delta_{j}\sigma_{i}\sigma_{j}+
(1-\delta_{i})(1-\delta_{j})S_{i}S_{j}+
\delta_{i}(1-\delta_{j})\sigma_{i}S_{j}+
(1-\delta_{i})\delta_{j}S_{i}\sigma_{j}\right]-H(t)\sum_{i}\left(\delta_{i}\sigma_{i}+(1-\delta_{i})S_{i}\right),
\end{equation}
where $J>0$ is the ferromagnetic exchange interaction energy between $i$ and $j$ sites. The $\sigma$ and $S$ are
conventional Ising spin variables which can take values of $\sigma=\pm 1/2$ and $S=\pm1, 0$ for the magnetic
$A$- and $B$- components of system, respectively. $\delta_{i}$ is a random variable which can take value of unity
or zero depending on whether the site-$i$ is occupied by  $A$ or $B$ ion, respectively.
The first summation in Eq. (\ref{eq1}) is over the nearest-neighbor  pairs
while the second one is over all lattice sites in the system. $H(t)$ denotes the time dependent oscillating
magnetic field  described as $H(t)=h_{0}sin(\omega t)$, where $t$ is time, $h_{0}$ and $\omega$
are amplitude and angular frequency of the driving magnetic field,
respectively. The period of the oscillating magnetic field is given by $\tau=2\pi/\omega$.

We simulate the system specified by the Hamiltonian in Eq. (\ref{eq1}) on
a $L \times L$, where $L=128$, square lattice under periodic boundary
conditions applied in all directions by benefiting from MC simulation with
single-spin flip Metropolis algorithm \cite{Binder, Newman}. The general
simulation procedure we use in this study is as follows: The simulation
begins from high temperature using random initial conditions, and then the system
is slowly cooled down with the reduced temperature steps $k_{B}\Delta T/J=0.015$,
where configurations were generated by selecting the sites randomly  through the lattice
and making single-spin-flip attempts,  which were accepted or rejected
according to the Metropolis algorithm.  The numerical data were generated over
50 independent sample realizations by running  the simulations
for 20000 MC steps per site after discarding the first 10000 steps.
This amount of transient steps is found to be sufficient for
thermalization for the whole range of the parameter sets.

Our program calculates the instantaneous values of the
magnetizations $M_{A}(t)$ and $M_{B}(t)$, and also the total
magnetization $M_{T}(t)$ at time $t$  as follows:

\begin{equation}\label{eq2}
M_{A}(t)=\frac{1}{L^2}\sum_{i=1}^{N_{A}} \sigma_{i},\quad \quad \quad
M_{B}(t)=\frac{1}{L^2}\sum_{i=1}^{N_{B}} S_{i},\quad \quad \quad M_{T}(t)=M_{A}(t)+M_{B}(t),
\end{equation}
where $N_{A}$ denotes the total number of $A$ ions ($N_{A}=pL^2$) while $N_{B}$
represents the total number of $B$ ions ($N_{B}=(1-p)L^2$). Further, utilizing the instantaneous
magnetizations stated in Eq. (\ref{eq2}), we determine the dynamic loop area which measures the
energy loss of the system due to the hysteresis, and dynamic correlations between time dependent
magnetic field and magnetizations,   respectively as follows \cite{Acharyya5}:

\begin{equation}\label{eq3}
A_{\alpha}=-\oint M_{\alpha}(t)dH, \quad \quad \quad C_{\alpha}=\frac{1}{\tau}\oint M_{\alpha}(t)H(t)dt,
\end{equation}
here $\alpha= A, B$ and $T$. Namely, $A_{A}, A_{B}$, and $A_{T}$ correspond to the dynamic loop areas while
$C_{A}, C_{B}$, and $C_{T}$  refer to the dynamic correlations of magnetic  components $A$, $B$ and
total system, respectively.

In order to specify the location of dynamic phase transition point at which
\textbf{F} and \textbf{P} phases separate from each other, we use and
check the thermal variation of dynamic heat capacity and which is defined as:

\begin{equation}\label{eq4}
    C=\frac{dE_{Coop}}{dT},
\end{equation}
where $T$ is temperature while $E_{Coop}$ is the  cooperative part
of the energy over a full cycle of the external applied
magnetic field which is defined according
to the following equation:
\begin{equation}\label{eq5}
E_{Coop}=-\frac{1}{\tau L^2}\oint\left[
J\sum_{\langle i,j \rangle}\left[\delta_{i}\delta_{j}\sigma_{i}\sigma_{j}+
(1-\delta_{i})(1-\delta_{j})S_{i}S_{j}+
\delta_{i}(1-\delta_{j})\sigma_{i}S_{j}+
(1-\delta_{i})\delta_{j}S_{i}\sigma_{j}\right]\right]dt.
\end{equation}

\section{Results and Discussion}\label{discussion}

In this section, in order to explain the dynamic evolution
of the magnetic system in detail, we will focus our attention on non-equilibrium
phase transition properties as well as stationary-state treatment
of the quenched disordered binary  ferromagnetic alloy system under
a time dependent oscillating magnetic  field. We argue and discuss
how the amplitude and period of the external oscillating magnetic
field as well as concentration of $A$-type magnetic component affect the
dynamic critical nature of the system. We will also shed
light on the physical facts lying  behind the thermal dependencies
of dynamic loop areas as well as correlations between time dependent
magnetizations and magnetic field for some selected combination of
system parameters.  As a final investigation, we examine the  variations of coercivity and
also hysteresis behavior as functions of  system parameters, as well as
active concentration of $A$-type magnetic component.

\begin{figure}[!here]
\begin{center}
\includegraphics[width=14.0cm,height=7.5cm]{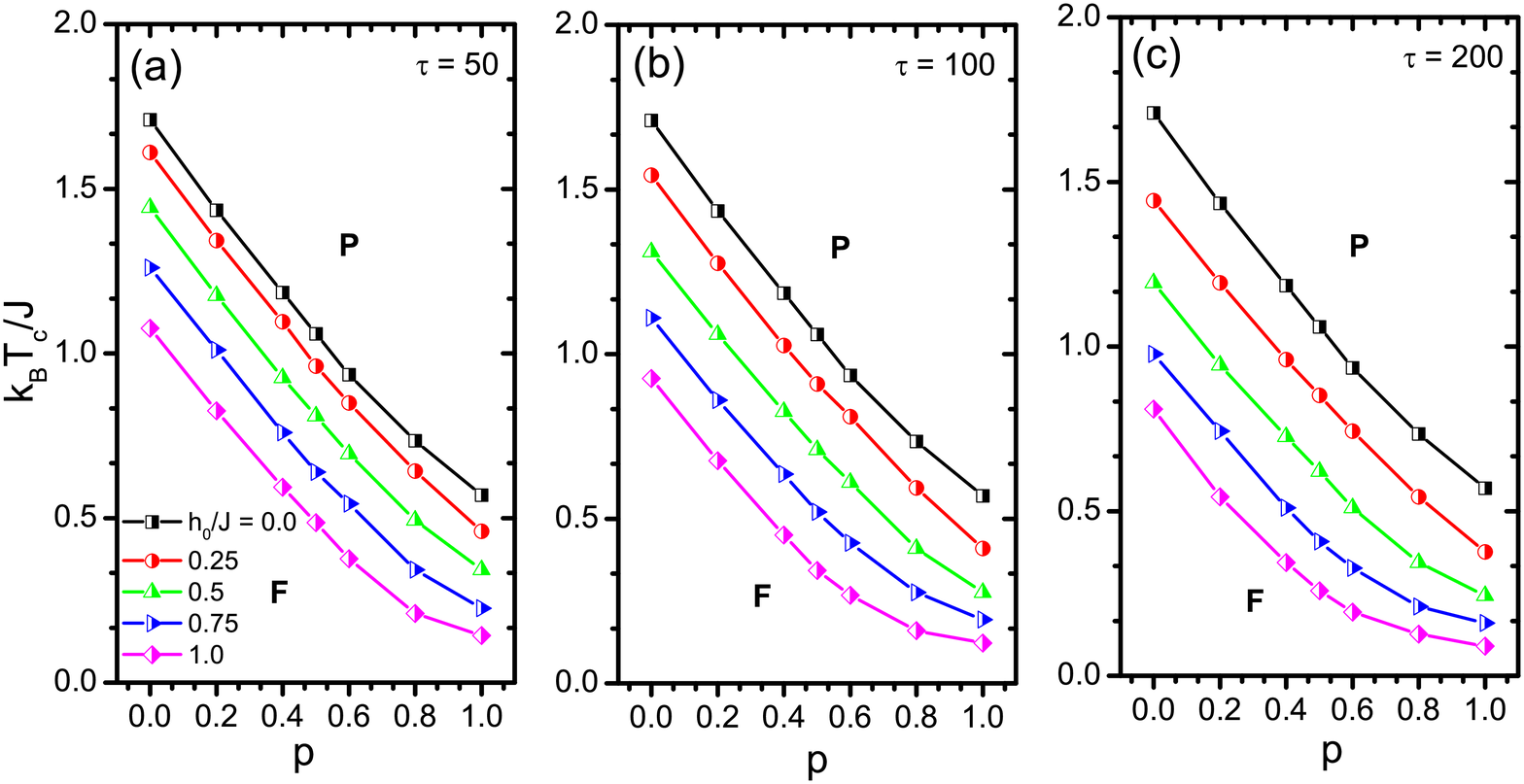}
\caption{Dynamic phase diagrams in $(p-k_{B}T_{C}/J)$ planes obtained from the peaks of
the dynamic heat capacity for values of $h_{0}/J=0.0, 0.25, 0.5, 0.75$ and 1.0.
The curves are demonstrated for three values of oscillation periods: (a) $\tau=50$,
(b) $\tau=100$ and (c) $\tau=200$. The characters \textbf{F} and \textbf{P} refer to the dynamically
ferromagnetic and paramagnetic phases, respectively.}\label{Fig2}
\end{center}
\end{figure}

Before going further, it should be  emphasized that for  $h_{0}/J=0.0$, i.e. static case, our numeric MC simulation
findings are  completely in accordance with the recently published
work \cite{Cambui} where thermal equilibrium phase transition properties of
the system we consider have been addressed by utilizing MC simulation with Metropolis algorithm.
Let us start to discuss the non-equilibrium phase transition properties of the system.
In Figs. \ref{Fig2}(a-c), we give  the global dynamic phase boundaries that distinguish
between \textbf{F} and \textbf{P} phases in a ($p-k_{B}T_{c}/J$) space with
five reduced external field amplitude values,
such as $h_{0}/J=0.0, 0.25, 0.5, 0.75$ and $1.0$ for some considered values of oscillation
periods, namely $\tau=50, 100$ and $200$.  At first sight, one can easily see
from the figure that, $(i)$ for a fixed set of values of $h_{0}/J$ and $\tau$, when the
concentration of spin-$1/2$ atom is increased starting from $p= 0.0$ \cite{note1},
\textbf{F} to \textbf{P} phase transition point moves to a lower value in
temperature axis. This is because of the fact that  the system tends to become disordered due to
the occurrence of two different types of randomly occupied magnetic
components as the active concentration of spin-$1/2$
is increased starting from $p = 0.0$.

It is possible to say that as the $p$ value is increased then the energy contribution
which comes from spin-spin interactions gets smaller. Consequently,
the \textbf{F} to \textbf{P} phase transition point moves to a lower
temperature, due to the energy resulting from the temperature and (or) magnetic
field which overcome the ferromagnetic spin-spin interactions.
Therefore, the \textbf{F} regions in $(p-k_{B}T_{C}/J)$ plane get narrower.
Moreover, $(ii)$ for a fixed set of values of $p$ and $\tau$, we see
that when $h_{0}/J$ increases, then the magnetic energy
supplied by the external applied field dominates against
the ferromagnetic spin-spin exchange interactions, and hereby,
the studied system can relax within the oscillation period $\tau$
of the external field which gives rise to a reduction in the
transition temperature. In order to see this physical fact, one can
compare any two values of  applied field amplitudes,
for example  $h_{0}/J=0.25$ with 0.5  for $p=0.8$
in Fig.\ref{Fig2}(a). Another important finding of
our numerical simulation is that the location of the
phase transition temperature sensitively depends on
the applied field period. $(iii)$  For a  fixed set
of values $p$ and $h_{0}/J$, it is obvious that as
$\tau$ is increased, the phase transition point is lowered
since decreasing field frequency gives rise to a decreasing phase
delay between the magnetizations and  magnetic field (i.e. the magnetizations
can  follow the oscillating driving magnetic field)
and this makes the occurrence of the dynamic phase transition easy.
The physical discussions mentioned above can be easily seen by checking any
two values of oscillation periods such as $\tau=100$ and $200$ for fixed
values of $h_{0}/J=0.5$ and $p=0.8$.

\begin{figure}[!here]
\begin{center}
\includegraphics[width=14.0cm,height=7.5cm]{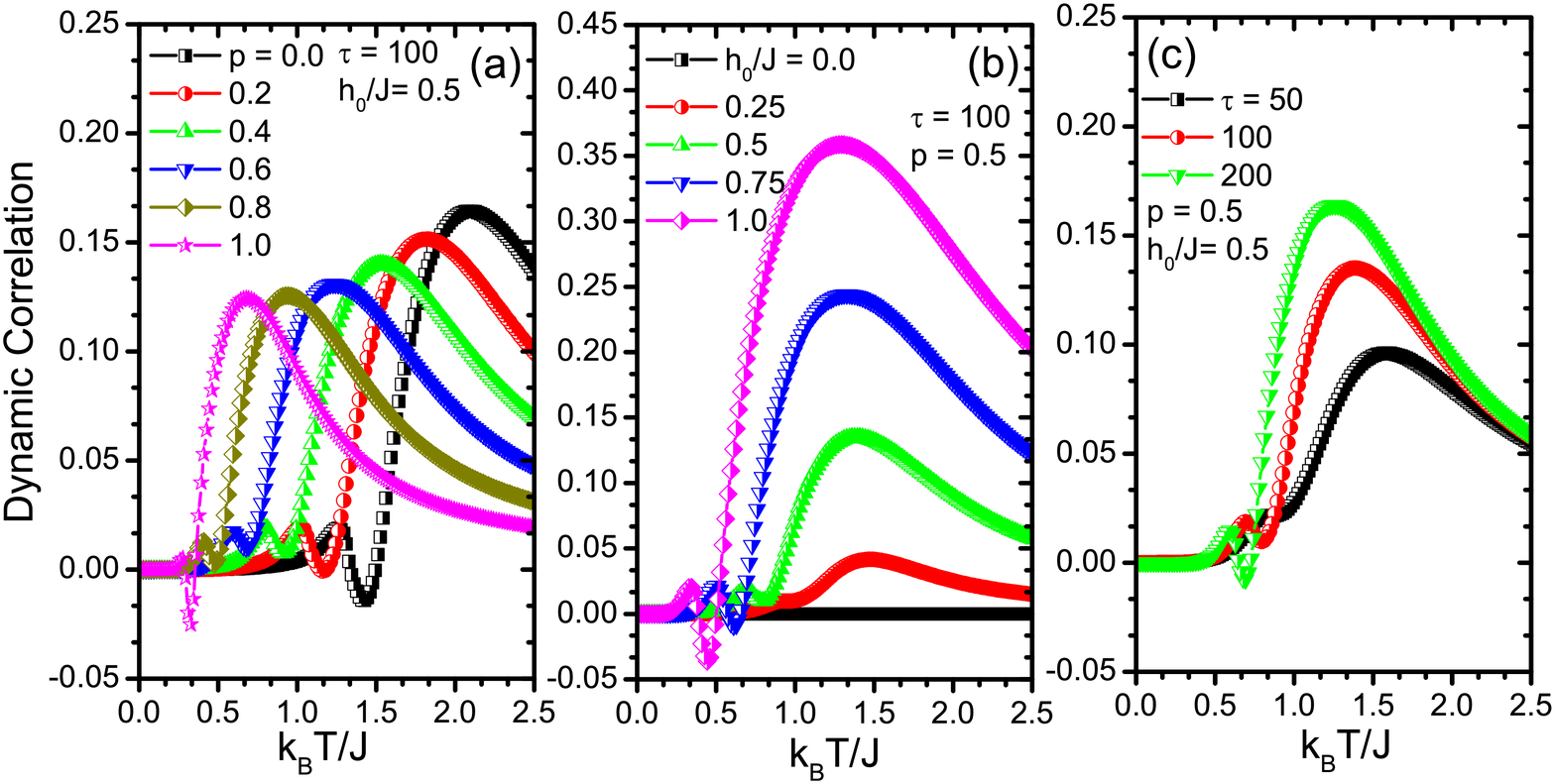}
\caption{Thermal variations of  dynamic total correlations corresponding to
the dynamic phase diagrams shown in Figs. \ref{Fig2}(a-c). The parameters we study
are remarked on the graphs.}\label{Fig3}
\end{center}
\end{figure}

We draw in Figs. \ref{Fig3}(a-c) the active concentration of $A$-type magnetic component,
reduced amplitude and oscillation period of field dependencies of dynamic correlation versus
temperature curves. One can able to deduce from the figures that, at
relatively low temperature regions where the system exists in \textbf{F} phase,
the dynamic correlations between time dependent magnetizations of the
system and  forcing field are close to zero
since  thermal energy is almost negligible, and the ferromagnetic spin-spin
interactions are dominant against the field energy. Moreover, the system
begins to show a shallow dip behavior, which may become negative depending on the
system parameters,  in the vicinity of dynamic phase transition point of system.
For example, the dynamic correlation curves exhibit concentration,
reduced amplitude and oscillating period of field induced
negative shallow dip behavior in Figs. \ref{Fig3}(a), (b) and (c), respectively.

\begin{figure}[!here]
\begin{center}
\includegraphics[width=14.0cm,height=7.5cm]{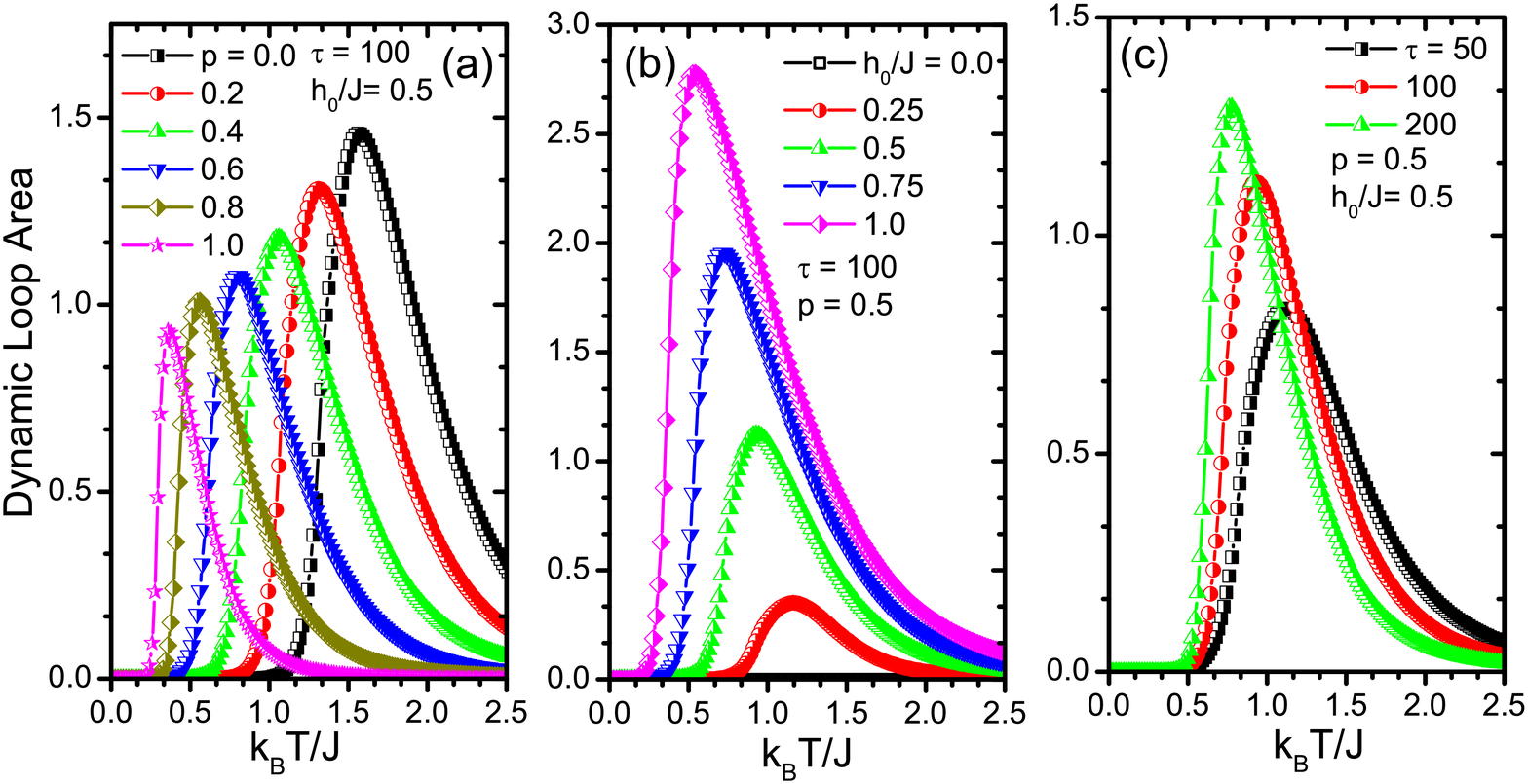}
\caption{Temperature variations of  dynamic total loop areas for varying system parameters which are
indicated in related graphs. }\label{Fig4}
\end{center}
\end{figure}

\begin{figure}[!here]
\begin{center}
\includegraphics[width=9.5cm,height=7.5cm]{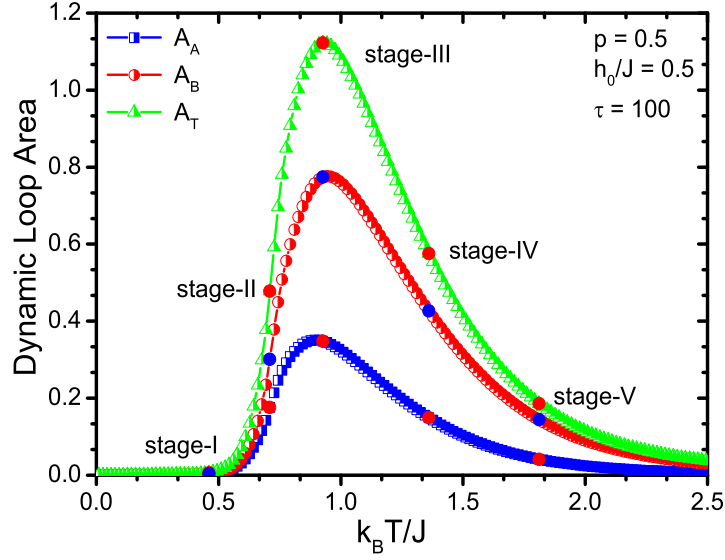}
\caption{Thermal variations of dynamic loop areas of magnetic
components $A$, $B$ and overall  system for considered values of system
parameters, such as $p=0.5, h_{0}/J=0.5$ and $\tau=100$.
The labeled stage-$i$ ($i=$I, II,..,V)  refers  to the five
different temperature locations indicated in graph.}\label{Fig5}
\end{center}
\end{figure}

In the following analysis, in order to show what happens in dynamic loop area versus temperature curves
at varying system parameters, we give in Figs. \ref{Fig4}(a-c) thermal variations of dynamic
loop areas corresponding to the dynamic phase diagrams illustrated in Figs. \ref{Fig2}(a-c).
As indicated in section (\ref{formulation}), dynamic loop area characterizes energy
dissipation due to the hysteresis. It can be  readily seen from the figures that the dynamic
loop area curves  obtained for a wide range of system parameters reveal a smooth and
rounded cusp located at a certain temperature above the dynamic phase
transition temperature of system. It would be interesting to see the origin of the physical
mechanisms underlying on the thermal treatment of dynamic loop areas.
Keeping this in mind, we separated the temperature space into five
stages which are shown in Fig. \ref{Fig5} where dynamic loop areas of two different parts as
well as of overall system have been plotted for selected values of system
parameters such as $p=0.5,\; h_{0}/J=0.5$ and $\tau=100$. At stage-I, the temperature
is sufficiently low so that both magnetic  components $A$ and $B$ of the
system show a dynamic ferromagnetism. In other words,   the time dependent
magnetizations of the system can not follow  the external field simultaneously.
Hence, dynamic loop areas of the system are  close to zero.
If we look at stage-II, where the location of the stage-II corresponds to the
dynamic critical temperature of the system for the parameters
we select, we can see that an increment  in temperature
gives rise to a reduction of ferromagnetism, and
time dependent magnetizations of the system begin to relatively
respond to the time dependent external magnetic field. As a result of this,
hysteresis loop areas of both magnetic components $A$ and $B$ of the
system get wider. Before going further, we note that even though
the stage-II is a phase transition point, the dynamic loop areas does
not show a maximum peak at the position of stage-II. This is because of such  MC computer experiments include of
thermal fluctuations of spins. If one considers the same problem with MFT (or EFT)
which neglects (or partially includes) the thermal fluctuations of spins, and focuses
on the thermal evolution of dynamic loop areas of the system, with a high probability,
one can see that thermal variation of dynamic loop area  may exhibit a maximum at
the dynamic phase transition point \cite{Vatansever2, Acharyya5}.
Let us continue with stage-III where the dynamic loop areas of the system present
a maximum peak, both magnetic components $A$ and $B$ of
the system show a dynamic paramagnetism, i.e., the time dependent
magnetizations of the system can follow the external field with a relatively small
phase lag. As the temperature is increased above the dynamic transition
point,  we reach the stages-IV and -V, where the value of dynamic
loop areas begin to reduce, respectively. Apart from the changing values of
the dynamic loop area corresponding to different stages we determine,
shapes of hysteresis curves sensitively depend on the selected system parameters
which will be discussed below.

\begin{figure}[!here]
\begin{center}
\includegraphics[width=14.0cm,height=7.5cm]{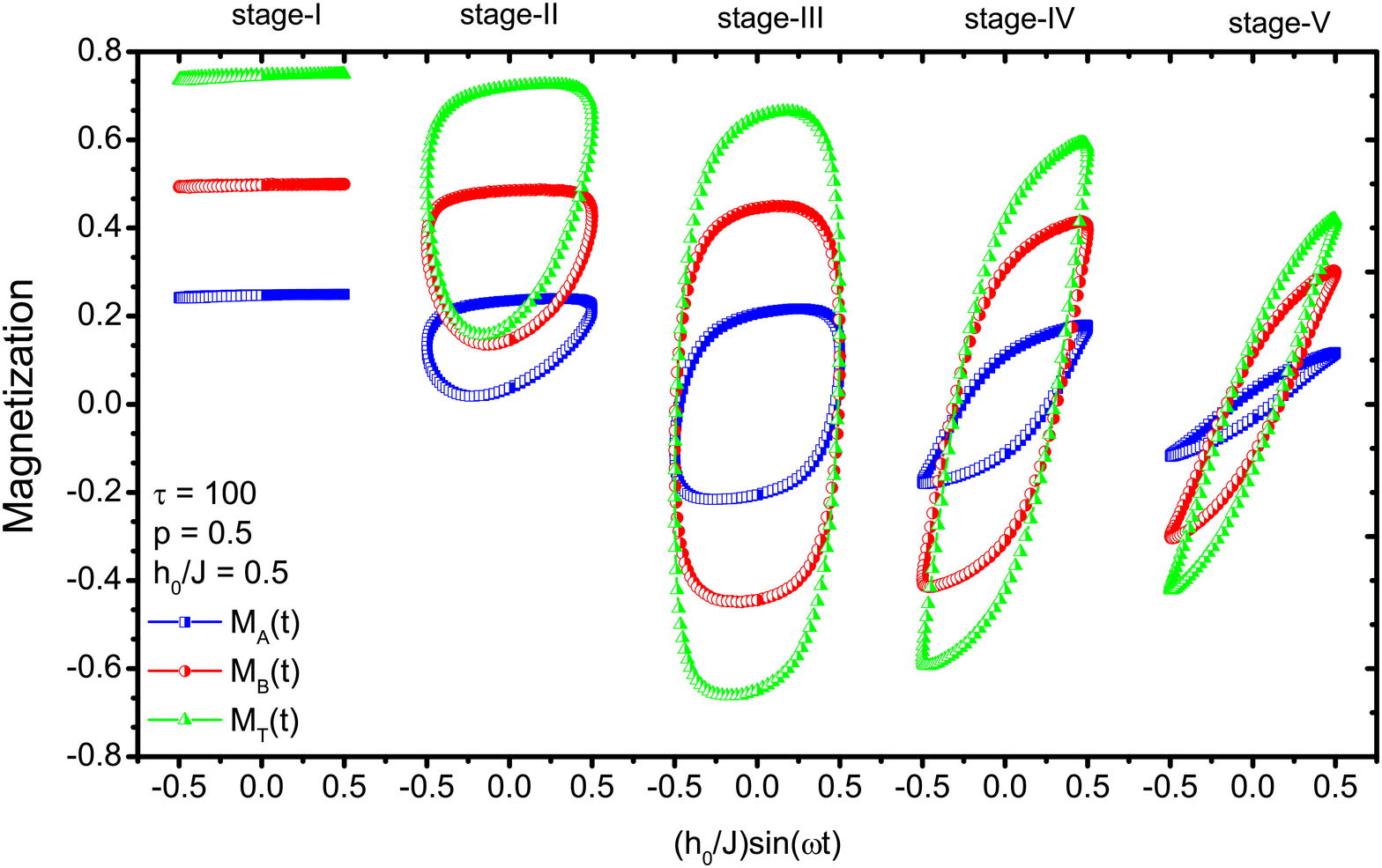}
\caption{Hysteresis curve evolutions of  both magnetic
components $A$ and $B$ and also overall of system at different five temperature
locations for the same system parameters with figure \ref{Fig5}. }\label{Fig6}
\end{center}
\end{figure}

In order to understand the varying temperature influences on the hysteresis
treatments of the system, we give in Fig. \ref{Fig6} the hysteresis curves
for both magnetic components $A$, $B$ and overall of the system for selected system parameters
corresponding to the diagram depicted in Fig. \ref{Fig5}.
When we look at the hysteresis curves obtained for stage-I,
we see that hysteresis loops are asymmetric around zero value
indicating an \textbf{F} phase. If one moves throughout stage-I $\rightarrow$ stage-II
way,  it can be easily seen that the
asymmetric shapes of  hysteresis curves begin to slowly break down owing to
the existence of a higher thermal energy than before.  When the studied system attains the stage-III,
it is possible to observe an example of square-like shape of hysteresis curves.
In this zone, the system completely exhibits a \textbf{P} character.
As the temperature increased, we reach the  stage-IV, and we see that shape of
hysteresis loops begin to change from square-like to sigmoidal-like  shaped due to
the thermal agitations. In addition to these, further increment in temperature
leads to the occurrence of a narrower sigmoidal-like hysteresis
curves  which are explicitly shown in stage-V of figure. Similar 
of observations have been recently found in ultrathin
Blume-Capel films under a magnetic field which oscillates in
time \cite{Yuksel3}.

\begin{figure}[!here]
\begin{center}
\includegraphics[width=14.0cm,height=7.5cm]{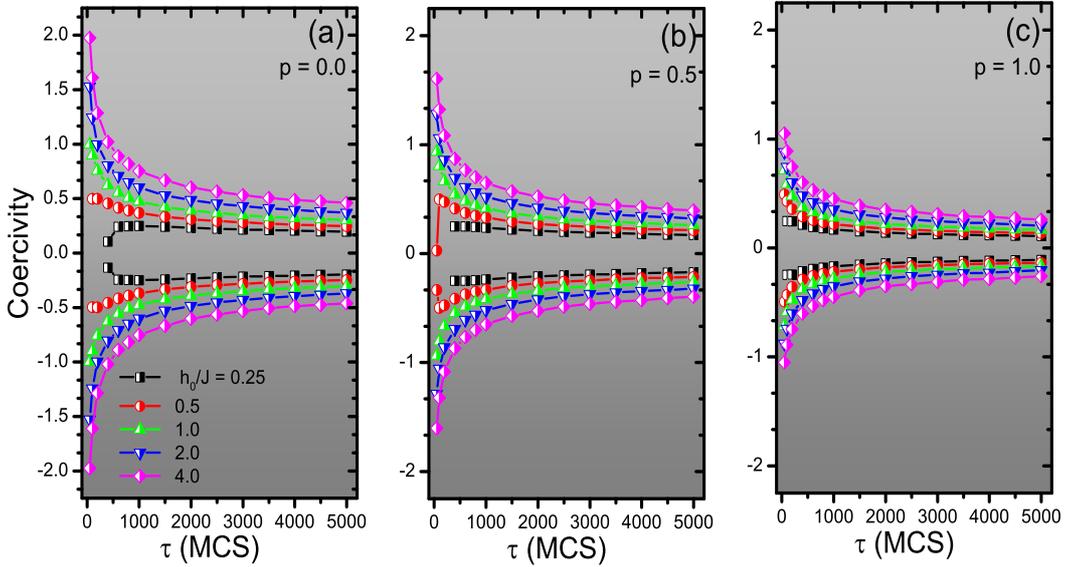}
\caption{Variations of the coercive fields as
a function of the oscillation period for several values of $h_{0}/J$
with (a) $p=0.0$, (b) $p=0.5$ and (c) $p=1.0$, respectively.}\label{Fig7}
\end{center}
\end{figure}

\begin{figure}[!here]
\begin{center}
\includegraphics[width=14.0cm,height=7.5cm]{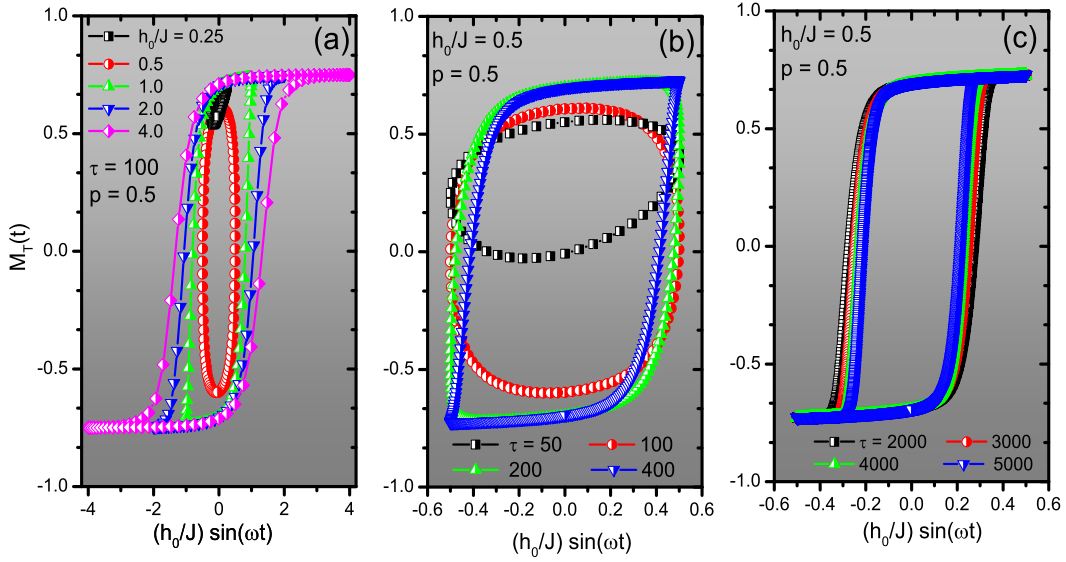}
\caption{Influences of (a) reduced amplitudes for $\tau=100$, (b)  relatively low periods, and
(c)  high periods of field for $h_{0}/J=0.5$ on the  hysteresis loops of the system corresponding to the
figure \ref{Fig7}(b).}\label{Fig8}
\end{center}
\end{figure}

In the following two analysis, namely in Figs. \ref{Fig7} and \ref{Fig8},
we investigate the oscillation period,  reduced amplitude of external
applied field and concentration of $A$-type magnetic component dependencies
of the coercivity as well as hysteresis curves of the system.
The aforementioned physical quantities have been calculated at a
temperature $T=0.8T_{c}^{0}(p)$, where $T_{c}^{0}(p)$ is the static phase
transition point in the absence of the external field, and it depends on the
studied active concentrations of magnetic atoms. The reason why we
choose such a temperature is that the system can be capable of undergoing a 
purely mechanical phase transition, such as applied field period or reduced
amplitude induced phase transition. The numerical data were collected for 100 cycles
of the external field after discarding the first 100 cycles of field have been discarded to
obtain a stationary state behavior.

We give the period of external field dependencies of the coercivity of the
system in Fig. (\ref{Fig7}) with varying values of reduced applied field amplitudes, such
as $h_{0}/J=0.25, 0.5, 1.0, 2.0$ and $4.0$ with three considered values of
active concentration of $A$-type magnetic  component, namely $p=0.0$, $0.5$ and $1.0$.
It can be said by focusing on Fig. \ref{Fig7}(a) that
depending on the reduced applied field  amplitude,  it is possible to observe a
large coercive field at low applied field periods.  Moreover,  it is found
that coercivity curves of system exhibit a sudden variation with
increasing applied  field period whereas for sufficiently high $\tau$ they exhibit
a stable profile. Another important point we want to emphasize is that the
system has no coercivity value in the case of low values of applied field
period and reduced amplitude, for example $\tau=100$ and $h_{0}/J=0.25$. In this regime,
the system exists in \textbf{F} phase. However,  with a significant
increment in value of amplitude for the same period of field causes the existence of a
purely mechanical reduced amplitude induced  phase transition, and coercivity treatment begins
to show itself naturally. The nearly same discussions are also  valid for Figs \ref{Fig7}(b-c)
which are plotted for $p=0.5$ and $p=1.0$, respectively. As the type-$A$ magnetic
component is introduced into the lattice, the coercivity value decreases prominently
since the energy originating from the time dependent forcing field overcomes the ferromagnetic
spin-spin interaction term between spin pairs in the lattice. Moreover, we should
note that  at the relatively low values of the amplitude and the period of the external
field, even though some small number of spin flip occurs, the considered alloy system shows
a dynamic ferromagnetic character. The physics mentioned briefly here is valid for all
concentration values of the system.

As a final investigation, we present the evolution of some typical
hysteresis curves corresponding to the coercivity properties illustrated
in Fig. \ref{Fig7}. Throughout Figs. \ref{Fig8}(a-c), we fix the active concentration
of magnetic components as $p=0.5$ and change the external field
components, namely period and reduced amplitude of field.
In Fig. \ref{Fig8}(a),  we select the period of field as $\tau=100$ to investigate the influences of
varying reduced amplitudes of field on hysteresis curves. It is clear from the figure that
while the system exists in dynamically  ordered phase and hysteresis curve is  asymmetric around
zero value at the relatively low reduced external  field amplitude, i.e. $h_{0}/J=0.25$, with increasing
amount of  $h_{0}/J$ the system starts to show a dynamically paramagnetic character,
and hysteresis curve is symmetric around zero value. According to our numerical findings,
another important  observation is that the shape of the hysteresis curves sensitively depend on the
applied field amplitude, for example, while the hysteresis loop is broad square-like
shaped for value of $h_{0}/J=0.5$, it is sigmoidal-like shaped for values
of $h_{0}/J=1.0$, $2.0$ and $4.0$. In Fig. \ref{Fig8}(b), we give the
dynamic hysteresis loops obtained for selected values of applied field period
such as $\tau=50, 100, 200$ and $400$ with fixed value of $h_{0}/J=0.5$.
At first glance, by comparing the curves obtained for $\tau=50$ and $100$, one can
clearly deduce from the figures that \emph{(i)} remanent magnetization  and also coercivity values of the system tend to
change, \emph{(ii)} the hysteresis curves begin to widen indicating a period  of field induced phase
transition with increasing $\tau$. If we increase $\tau$,
the hysteresis curves becomes broad square-like shaped. On the other hand, in Fig. \ref{Fig8}(c), it is possible to see that
the hysteresis curves are  sigmoidal shaped, and the width of the loops becomes narrower but does not
vanish  at the relatively high values of  applied field period regimes. This type of behavior has been
reported in \cite{Yuksel2, Vatansever3} where non-equilibrium phase transition properties of a nanocube
system is investigated with MC method of simulation.

\section{Conclusion}\label{conclusion}
In conclusion, by making use of MC computer experiment method with Metropolis algorithm,
we have looked for answers  for the questions discussed in section \ref{introduction}.
As we mentioned above,  as a disordered binary ferromagnetic alloy with coupling $J$, we have selected a system
consisting of spin-$1/2$ and spin-$1$ components, where the spin-$1/2$ and spin-$1$ components are distributed randomly depending upon the controllable
concentration value. The situations of $p=1.0$ and $0.0$ correspond to the completely pure kinetic
Ising and Blume-Capel models without single-ion anisotropy, respectively.
Namely, for both two values of $p$ we remark, there is no lattice including disorder originating from different
species of components in the system. However, it is obvious that except from these two values of $p$,
the magnetic system composes of randomly distributed spin-$1/2$ and spin-$1$ components where the weights
of the components strongly depend on the concentration values. After  detailed numerical operations,
the most prominent findings underlined in the present paper
can be briefly summarized as follows:

\begin{itemize}
\item  For a fixed set of values of $h_{0}/J$ and $\tau$, we found an almost linear decrease
the critical temperature with increasing type-$A$ content.

\item We see that when $h_{0}/J$ increases, then the magnetic energy supplied by
the external applied field dominates against the ferromagnetic spin-spin exchange interactions,
and hereby, the studied system can relax within the oscillation period $\tau$ of the external
field which gives rise to a reduction in the transition temperature, for a fixed set of values $p$ and $\tau$.

\item  For a  fixed set of values $p$ and $h_{0}/J$, as $\tau$ is increased, the phase
transition point is lowered because of increasing field frequency gives rise to a growing phase delay between the time
dependent magnetizations and  forcing applied field.
\end{itemize}

\noindent The situations corresponding to the explanations we emphasize above can be easily seen by looking
the dynamic phase boundaries separating the \textbf{F} and \textbf{P} phases which have been constructed in
active concentration of spin-1/2 atoms and dynamic critical temperature planes at
various Hamiltonian parameters in Fig. \ref{Fig1}. Our simulation results also suggest that:

\begin{itemize}
\item  For studied values of system parameters,  thermal variations of dynamic correlations
exhibit period and reduced amplitude of field as well as  active concentration of $A$-type magnetic component
induced local dip behavior.
\item The locations of maximum lossy points which are characterized by hysteresis loop area
show a decreasing tendency in temperature value with increasing $p$ values starting from zero
for considered values of Hamiltonian parameters. Other system parameter dependencies of
dynamic loop area has also been  discussed in the present study.

\item Applied field period dependencies of the coercivities obtained for three
concentration values of $p$ with varying reduced amplitudes of field reveal that
they exhibit a stable profile at the relatively high period of field whereas
they show a sudden change in the low period of field. It is also found that for a fixed set
of $h_{0}/J$ and $\tau$, as the $p$ value increases starting from zero, the corresponding
coercivity value begins to decrease this is because of the amount of spin-1/2
atoms increases in system.
\item Shape of the hysteresis curve of the system strongly depends on the taking Hamiltonian
parameters into consideration. For example, while it is possible to observe an example of a
broad square-like shaped hysteresis curve at the relatively low period of field, the shape of it
begins to change to be a sigmoidal-like shape with increasing period of field.
\end{itemize}

All of the observations found in this work show that the amplitude and frequency of the
driving time dependent magnetic field as well as active concentration of $A$-type magnetic
component have an important influence on the thermal and magnetic properties, such as
coercivity and critical temperature of the system.

Finally, we note that such a binary alloy system can actually contain four different values of
relative $A-B$ interaction energy, namely $J_{AA}$, $J_{BB}$, $J_{AB}$, $J_{BA}$ corresponding
to the randomly distributed components between $A-A$, $B-B$, $A-B$ and $B-A$, respectively.
According to the findings of the previously published equilibrium phase transition studies mentioned in
section \ref{introduction}, disordered binary alloy system with different signs and
unequal magnitudes of the spin-spin interactions and single ion-anisotropy
reveals interesting thermal and magnetic behaviors such as existence of re-entrant,
compensation and also spin-glass behaviors. In this work, we have focused on only
the non-equilibrium phase transition properties of disordered binary alloy
system in cases of $J_{AA}=J_{BB}=J_{AB}=J_{BA}=J$ and also zero single-ion
anisotropy. So, determination of the non-equilibrium phase
transitions features of a disordered  binary alloy including different types of
spin-spin interactions and also single-ion anisotropy  may be an interesting study.
Furthermore,  based on the earlier  systems \cite{Punya, Vatansever2, Ozan2},
it is possible to categorize the frequency dispersions of dynamic loop areas into
three distinct types depending on the system parameters. Keeping in this mind, it would also be
very interesting to see the frequency  dispersions of dynamic loop areas of the same
system.

\section{Acknowledgements}
The numerical calculations reported in this paper were
performed at T\"{U}B\.{I}TAK ULAKB\.{I}M (Turkish agency), High Performance and
Grid Computing Center (TRUBA Resources).


\begin{thebibliography}{99}
\bibitem{Tome} T. Tom\'{e}, M.J. de Oliveira, Phys. Rev. A 41 (1990) 4251.
\bibitem{Suen} J.-S. Suen, J.L. Erskine,  Phys. Rev. Lett. 78  (1997) 3567.
\bibitem{He} Y.-L He, G.-C. Wang, Phys. Rev. Lett. 70   (1993) 2336.
\bibitem{Jiang} Q. Jiang, H.-N. Yang, G.-C. Wang, Phys. Rev. B 52 (1995) 14911.
\bibitem{Choi} B.C. Choi, W.Y. Lee, A. Samad, J.A.C. Bland, Phys. Rev. B 60 (1999) 11906.
\bibitem{Robb} D.T. Robb, Y.H. Xu, O. Hellwig, J. McCord, A. Berger, M.A. Novotny,
P.A. Rikvold, Phys. Rev. B 78 (2008) 134422.
\bibitem{Berger} A. Berger, O. Idigoras, P. Vavassori, Phys. Rev. Lett. 111  (2013) 190602.
\bibitem{Acharyya1} M. Acharyya, Phys. A 253  (1998) 199.
\bibitem{Buendia} G.M. Buend\'{i}a, M. Machado, Phys. Rev. E 58  (1998) 1260.
\bibitem{Punya} A. Punya, R. Yimnirun, P. Laoratanakul, Y. Laosiritaworn, Physica B 405 (2010) 3482.
\bibitem{Keskin} M. Keskin, O. Canko, U. Temizer, Phys. Rev. B 72  (2005) 036125.
\bibitem{Keskin2} M. Keskin, E. Kantar, O. Canko, Phys. Rev. E 77  (2008) 051130.
\bibitem{Idigoras} O. Idigoras, P. Vavassori, A. Berger, Physica B 407  (2012) 1377.
\bibitem{Shi} X. Shi, G. Wei, L. Li, Phys. Lett. A 372  (2008) 5922.
\bibitem{Shi2} X. Shi, J. Zhao, X. Xu, Physica A, 419  (2015) 234.
\bibitem{Yuksel} Y. Yuksel, E. Vatansever, U. Akinci, H. Polat, Phys. Rev. E 85  (2012) 051123.
\bibitem{Akinci} U. Akinci, Y. Yuksel, E. Vatansever, H. Polat, Physica A 391  (2012) 5810.
\bibitem{Vatansever1} E. Vatansever, B.O. Aktas, Y. Yuksel, U. Akinci, H. Polat, J. Stat. Phys. 147 (2012) 1068.
\bibitem{Vatansever2} E. Vatansever, U. Akinci, Y. Yuksel, H. Polat, J. Magn. Magn. Mater. 329  (2013) 14.
\bibitem{Ozan} B.O. Aktas, U. Akinci, H. Polat, Phys. Rev. E 90  (2014) 012129.
\bibitem{Shi3} X. Shi, G. Wei, Phys. Scr. 89  (2014) 075805.
\bibitem{Lo} W.S. Lo, R.A. Pelcovits, Phys. Rev. A 42  (1990) 7471.
\bibitem{Chatterjee} A. Chatterjee, B.K. Chakrabarti, Phys. Rev. E 67  (2003) 046113.
\bibitem{Park} H. Park, M. Pleimling, Phys. Rev. E 87  (2013) 032145.
\bibitem{Acharyya2} M. Acharyya, B.K. Chakrabarti, Phys. Rev. B 52 (1995) 6550.
\bibitem{Acharyya3} M. Acharyya, J. Magn. Magn. Mater. 323  (2011) 2872.
\bibitem{Acharyya4} M. Acharyya, Phys. Rev. E 69  (2004) 027105.
\bibitem{Acharyya5} M. Acharyya, Phys. Rev. E 58 (1998) 179.
\bibitem{Tauscher} K. Tauscher, M. Pleimling, Phys. Rev. E 89  (2014) 022121.
\bibitem{Park2} H. Park, M. Pleimling, Phys. Rev. Lett. 109  (2012) 175703.
\bibitem{Yuksel2} Y. Yuksel, E. Vatansever, H. Polat, J. Phys.: Condens. Matter 24  (2012) 436004.
\bibitem{Vatansever3} E. Vatansever, H. Polat, Physica A 394  (2014) 82.
\bibitem{Vatansever4} E. Vatansever, H. Polat, arXiv:1311.3537v2.
\bibitem{Zheng} G.-P. Zheng, M. Li, Phys. Rev. B 66 (2002) 054406.
\bibitem{Ozan2} B.O. Aktas, U. Akinci, H. Polat, Physica B 407 (2012) 4721.
\bibitem{Vatansever5} E. Vatansever, U. Akinci,  H. Polat, J. Magn. Magn. Mater. 344   (2013) 89.



\bibitem{Thorpe} M.F. Thorpe, A.R. McGurn, Phys. Rev. B 20   (1979) 2142.
\bibitem{Tahir} R.A. Tahir-Kheli, T. Kawasaki, J. Phys. C 10 (1977) 2207.
\bibitem{Plascak} J.A. Plascak, Physica A 198  (1993) 655.
\bibitem{Katsura} S. Katsura, F. Matsubara, Canad. J. Phys. 52  (1974) 120.
\bibitem{Ishikawa} T. Ishikawa, T. Oguchi,  J. Phys. Soc. Jpn. 44  (1978) 1097.
\bibitem{Honmura1}   R. Honmura, A.F. Khater, I.P. Fittipaldi, T. Kaneyoshi,  Solid State  Commun. 41 (1982) 385.
\bibitem{Kaneyoshi1} T. Kaneyoshi, Phys. Rev. B  34  (1986) 7866.
\bibitem{Kaneyoshi2} T. Kaneyoshi, Phys. Rev. B  33  (1986) 7688.
\bibitem{Kaneyoshi3} T. Kaneyoshi, Z.Y. Li,  Phys. Rev. B  35  (1987) 1869.
\bibitem{Kaneyoshi4} T. Kaneyoshi, Phys. Rev. B  39 (1989) 12134.
\bibitem{Kaneyoshi5} T. Kaneyoshi, J. Phys. Condens. Matter 5 (1993) L501.
\bibitem{Kaneyoshi6} T. Kaneyoshi, M. Ja\v{s}\v{c}ur,  J. Phys. Condens.  Matter 5  (1993) 3253.
\bibitem{Tatsumi} T. Tatsumi, Prog. Theor. Phys. 59 (1978) 1428; 59  (1978) 1437.
\bibitem{Scholten} P.D. Scholten,  Phys. Rev. B 32 (1985) 345.
\bibitem{Scholten2} P.D. Scholten,  Phys. Rev. B 40  (1989) 4981.
\bibitem{Godoy} M. Godoy, W. Figueiredo,  Int. J. Mod. Phys. C 20  (2009) 47.
\bibitem{Cambui} D.S. Cambui, A.S. De Arruda, M. Godoy,  Int. J. Mod. Phys. C 23  (2012) 1240015.
\bibitem{Binder}  K. Binder, Monte Carlo Methods in Statistical Physics (Springer, Berlin, 1979).
\bibitem{Newman} M.E.J. Newman, G.T. Barkema, Monte Carlo Methods in Statistical Physics (Oxford University Press, New York, 1999).
\bibitem{note1}The special cases of $p=1.0$ and $0.0$ indicate the pure spin-1/2 Ising and spin-1
Blume-Capel  without single-ion anisotropy models, respectively. This means that when concentration
of type-$A$ atom is selected to be as $p=1.0$, there is no any lattice site
occupied by a type-$B$ atom, the overall system composes of type-$A$ atom in the system, or vice versa.
\bibitem{Yuksel3} Y. Yuksel, Phys. Lett. A 377 (2013) 2494.
\end{thebibliography}
\end{document}